\documentclass[12pt]{article}
\usepackage[dvips]{graphicx}
\usepackage{amssymb, amsmath, mathbbol}
\usepackage[centerlast,footnotesize]{caption2}

\newcommand{\bra}{\bigl\langle}
\newcommand{\ket}{\bigr\rangle}

\newcommand{\Psibar}{\bar{\Psi}}
\newcommand{\psibar}{\bar{\psi}}

\newcommand{\Tr}{\operatorname{Tr}}
\newcommand{\lapprox}{\raisebox{-0.5ex}{$\
\stackrel{\textstyle<}{\textstyle\sim}\ $}}
\newcommand{\gapprox}{\raisebox{-0.5ex}{$\
\stackrel{\textstyle>}{\textstyle\sim}\ $}}

\usepackage{graphicx}
\newcommand{\One}{1\kern-4.5pt1}

\newcommand{\be}{\begin{equation}}
\newcommand{\ee}{\end{equation}}

\def\lesim{${\lower 2pt\hbox{$\scriptstyle
<$}\atop\raise 4pt\hbox{$\scriptstyle\sim$}}$} 
\def\grsim{${\lower2pt\hbox{$\scriptstyle >$} \atop\raise4pt\hbox 
{$\scriptstyle\sim$}}$} 
\begin{document}
\begin{center}
\begin{flushright}
October 2007
\end{flushright}
\vskip 10mm
{\LARGE
Hadron Spectrum in a Two-Colour Baryon-Rich Medium
}
\vskip 0.3 cm
{\bf Simon Hands$^{a,b}$, Peter Sitch$^a$, and Jon-Ivar Skullerud$^{c,d}$}
\vskip 0.3 cm
$^a${\em Department of Physics, Swansea University,\\
Singleton Park, Swansea SA2 8PP, U.K.}\footnote{Permanent address}
\vskip 0.3 cm
$^b${\em Isaac Newton Institute for Mathematical Sciences, \\20 Clarkson Road,
Cambridge CB3 0EH, U.K.}
\vskip 0.3 cm
$^c${\em School of Mathematics, Trinity College, Dublin 2, Ireland.}
\vskip 0.3 cm
$^d${\em Department of Mathematical Physics, National University of Ireland, 
\\Maynooth, County Kildare, Ireland.}\footnote{Permanent address}
\end{center}

\noindent
{\bf Abstract:} 
The hadron spectrum of SU(2) lattice gauge theory with two flavours of Wilson
quark is studied on
an $8^{3} \times 16$ lattice using all-to-all propagators, 
with particular emphasis on the dependence on quark chemical potential $\mu$.
As $\mu$ is increased from zero the diquark
states with non-zero baryon number $B$ 
respond
as expected, while states with $B=0$ remain unaffected
until the onset of non-zero baryon density at $\mu=m_{\pi}/2$. 
Post onset the
pi-meson mass increases in accordance with chiral perturbation theory
while the rho becomes lighter. In the diquark
sector a Goldstone state associated with a superfluid ground
state can be identified. A further consequence of superfluidity is an
approximate degeneracy between mesons and baryons with the same spacetime and 
isospin quantum numbers. Finally we find tentative evidence for the binding of
states with kaon quantum numbers within the baryonic medium.

                                                                                

\section{Introduction}

At large baryon chemical potential $\mu_B$ the properties of QCD are expected to
change as the system moves from a confined nuclear matter phase to
a deconfined quark matter phase where the relevant degrees of freedom are quarks
and gluons. At low temperature {\em T} and high $\mu_B$, the attraction between
quarks is believed to be sufficient to promote diquark Cooper pairing leading to
a colour superconducting ground state. Weak-coupling techniques can be used at
asymptotic densities and have revealed a superconducting phase known as the
colour-flavour locked phase.  However as density is reduced towards
phenomenologically reasonable values, the precise nature of the ground state
appears very sensitive both to additional parameters such as isospin
chemical potential and strange quark mass, and also to the nature of the
non-perturbative assumptions made in the calculation. It seems natural to 
use Lattice QCD to investigate these issues, but unfortunately whilst the
lattice has been used very successfully to investigate QCD with $T>0$, the well
known ``Sign Problem'' has made progress for $\mu_B/T\gg1$ impossible. 

Orthodox simulation techniques can be applied, however, to the case of 
two colour QCD (QC$_2$D) with gauge group SU(2). Whilst this theory differs in
important ways from QCD, for instance in having bosonic baryons in the
spectrum, and in having a superfluid, rather than superconducting, 
ground state at large $\mu_B$, it remains the simplest gauge theory in which
a systematic non-perturbative treatment of a baryonic medium is possible.
Moreover, recent simulations \cite{Hands:2006ve} have provided evidence that
there exist two distinct forms of two color matter: the dilute Bose gas formed
from diquark bound states which forms at onset, ie. 
for $\mu_B>\mu_{Bo}=M_\pi$, in which
superfluidity arises via Bose-Einstein condensation of scalar diquarks, is
supplanted at larger densities by a deconfined ``quark matter'' phase in which a
system of degenerate quarks is disrupted by BCS condensation at the Fermi
surface. Studies in this regime may have qualitative or even
quantitative relevance for QCD quark matter, particularly in the non-Goldstone
sector. For instance, Sch\"afer
\cite{Schafer:2002ty} has stressed how the impact of instantons on the
excitation spectrum at high baryon density could be elucidated by lattice
simulations.

In this Letter we study the $\mu_B$-dependence of the hadron
spectrum in both meson and baryon sectors of QC$_2$D with $N_f=2$ flavours of 
Wilson quark, which we find to vary dramatically as the onset from vacuum to
a ground state with non-zero baryon density is traversed. 
We build on the pioneering work of the Hiroshima group~\cite{Muroya:2002ry}, 
who found that beyond onset the pion mass did not change
noticeably, but the rho meson became significantly lighter, so that the
level-ordering is reversed. 
Two baryon channels were also studied but no significant $\mu_B$-dependence 
found. This work was performed on small lattices, and only looked at
a few states. One of our aims is to improve and
update their results. We also study the nature of the Goldstone 
mode associated with superfluidity, as done for QC$_2$D with staggered lattice
fermions in \cite{Kogut:2001na}, and expose the specifically two colour
phenomenon of ``meson-baryon'' mixing in the superfluid state, whereby since
$B$ is no longer conserved, states interpolated by mesonic operators
$q\bar q$ and diquark operators $qq$ have identical quantum numbers and hence 
exhibit an approximate degeneracy. Finally, we make the first 
measurements of strange meson masses in a baryonic medium, and find tentative   
evidence for bound states of kaons in nuclei.

\section{Formulation}

The gauge-invariant lattice action with $N_f=2$ degenerate
fermion flavours is~\cite{Hands:2006ve}
\begin{equation}
S=\bar\psi_1M(\mu)\psi_1+\bar\psi_2M(\mu)\psi_2-\kappa
j(\bar\psi_1K\bar\psi_2^T-\psi_2^TK\psi_1),
\label{eq:action}
\end{equation}
with $M$ the conventional Wilson fermion matrix (with lattice spacing $a=1$)
\begin{equation}
M_{xy}(\mu)=\delta_{xy}-\kappa\sum_\nu\left[
(1-\gamma_\nu)e^{\mu\delta_{\nu0}}U_\nu(x)\delta_{y,x+\hat\nu}+
(1+\gamma_\nu)e^{-\mu\delta_{\nu0}}U_\nu^\dagger(y)\delta_{y,x-\hat\nu}
\right],
\end{equation}
$\kappa$ the hopping parameter, $\mu$ the quark chemical potential, and 
$j$ the strength of an SU(2)$_L\otimes$SU(2)$_R$-invariant
diquark source term needed to regularise IR
fluctuations in the superfluid phase, which should be extrapolated to zero to
reach the physical limit. The subscript on the fermion fields is a flavour
index. The anti-unitary operator 
$K=K^T\equiv C\gamma_5\tau_2$,
where $C\gamma_\mu C^{-1}=-\gamma_\mu^T=-\gamma_\mu^*$ and the Pauli matrix
$\tau_2$ acts on colour indices. A useful relation is
\begin{equation}
M^T(\mu)=-K\gamma_5M(-\mu)K\gamma_5.
\label{eq:MDsymmetry}
\end{equation}
The hadronic states examined in this paper are $q\bar q$ mesons and $qq$, $\bar
q\bar q$ diquark baryons and anti-baryons. In all cases we use local
interpolating operators of the form $\bar\psi(x)\Gamma\psi(x)$,
$\psi^T(x)K\Gamma\psi(x)$, $\bar\psi(x)K\bar\Gamma\bar\psi^T(x)$. 
The matrix
$\Gamma=\gamma_0\bar\Gamma^\dagger\gamma_0$
determines the spacetime quantum numbers of the hadron, with inclusion
of the $K$ factor ensuring that mesons and baryons with the same $\Gamma$
have the same $J^P$. In this letter we will focus on states with 
$\Gamma\in\{\One,\gamma_5,\gamma_j,i\gamma_5\gamma_j\}$ with $j=1,\ldots,3$
corresponding to
$J^P\in\{0^+,0^-,1^-,1^+\}$.

\subsection{Fermion Propagators}

The fermion action (\ref{eq:action}) can be written in the form
$\Psibar\mathcal{M}(\mu,j)\Psi$
where $\Psibar\equiv(\psibar_1,\psi_2^T,\psibar_2,\psi_1^T)$ and
$\Psi\equiv(\psi_1,\psibar_2^T,\psi_2,\psibar_1^T)^T$.
It has the form:%
\begin{equation}
\label{form_or_m}
\mathcal{M}=\begin{pmatrix} A&0\\
            0&\bar{A}\end{pmatrix}
\;\;\;\mbox{with}
\;\;\;A=\textstyle\frac{1}{2}\begin{pmatrix} M&-\kappa jK\\
                                  \kappa jK&-M^T\end{pmatrix},
\end{equation}
and $\bar A(j)=A(-j)$.
Now consider the propagator%
\begin{equation}
\label{gorkov_propagator}
%\begin{split}
\bra\Psi(x)\Psibar(y)\ket 
%&=
%
%\begin{pmatrix} 
%\bra\psi_1(x)\psibar_1(y)\ket&\bra\psi_1(x)\psi_2^T(y)\ket&0&0\\
%\bra\psibar_2^T(x)\psibar_1(y)\ket&\bra\psibar_2^T(x)\psi_2^T(y)\ket&0&0\\
%   0&0&\bra\psi_2(x)\psibar_2(y)\ket&\bra\psi_2(x)\psi_1^T(y)\ket\\
%   0&0&\bra\psibar_1^T(x)\psibar_2(y)\ket&\bra\psibar_1^T(x)\psi_1^T(y)\ket
%   \end{pmatrix}\\&
\equiv \begin{pmatrix}
             S_{11}&S_{12}&0&0\\
	     \bar{S}_{21}&\bar{S}_{22}&0&0\\
             0&0& S_{22}&S_{21}\\
	     0&0& \bar{S}_{12}&\bar{S}_{11}
        \end{pmatrix}
%\end{split}
\end{equation}
where the zero entries arise from the assumption of isospin symmetry.
This symmetry also implies that%
\begin{equation}
\label{prop_symmetries}
S_{22}=S_{11}\equiv S_N;\quad \bar{S}_{11}=\bar{S}_{22}\equiv\bar S_N;
\quad S_{21}=-S_{12}\equiv S_A;\quad
\bar{S}_{12}=-\bar{S}_{21}\equiv-\bar S_A,
\end{equation}
where the subscripts denote ``normal'' and ``anomalous'' propagation. Anomalous
propagation arises from particle-hole mixing in a superfluid ground state; on
a finite volume $S_A$ vanishes in the limit $j\to0$.

\subsection{Mesons}

The isovector ($I=1$) meson operators $M^1$
are given by $\psibar_1\Gamma\psi_2$, $\psibar_2\Gamma\psi_1$ and 
$(\psibar_1\Gamma\psi_1-\psibar_2\Gamma\psi_2)/\surd{2}$. 
The charged meson correlator is then
  \begin{equation}
    \begin{split}
      \label{isovector_meson}
      \bra M^1(x)M^{1\dagger}(y)\ket 
&= \bra\psibar_1(x)\Gamma\psi_2(x)\psibar_2(y)\bar{\Gamma}\psi_1(y)\ket\\
      &= -\Tr[S_N(y,x)\Gamma S_N(x,y)\bar{\Gamma}] 
+ \Tr[\bar{S}_A(y,x)\Gamma S_A(x,y)\bar{\Gamma}^T]
    \end{split}
  \end{equation}
For the neutral meson correlator, 
the ``disconnected'' parts made up from the product of two traces
cancel because of isospin symmetry. 
The connected parts are:
    \begin{equation}
      \begin{split}
      \label{neutral_isovector_meson}
      \bra M^1(x)M^{1\dag}(y)\ket &= 
\bra\psibar_1(x)\Gamma\psi_1(x)\psibar_1(y)\bar{\Gamma}\psi_1(y)\ket_c 
      -\bra\psibar_1(x)\Gamma\psi_1(x)\psibar_2(y)\bar{\Gamma}\psi_2(y)\ket_c\\ 
      &\quad\mspace{2mu}-\bra\psibar_2(x)\Gamma\psi_2(x)
\psibar_1(y)\bar{\Gamma}\psi_1(y)\ket_c
      +\bra\psibar_2(x)\Gamma\psi_2(x)\psibar_2(y)\bar{\Gamma}\psi_2(y)\ket_c\\
      &= -\Tr[S_N(y,x)\Gamma S_N(x,y)\bar{\Gamma}] 
+ \Tr[\bar{S}_A(y,x)\Gamma S_A(x,y)\bar{\Gamma}^T]
      \end{split}
    \end{equation}
and so the degeneracy between neutral and charged isovector mesons is manifest.
The isoscalar ($I=0$) 
meson $M^0=(\psibar_1\Gamma\psi_1+\psibar_2\Gamma\psi_2)/\surd{2}$
has the correlator
    \begin{equation}
      \label{isoscalar_meson}
      \begin{split}
      \bra M^0(x)M^{0\dag}(y)\ket = 2&\Tr[S_N(x,x)\Gamma]
\Tr[S_N(y,y)\bar{\Gamma}]\\
      -&\Tr[S_N(y,x)\Gamma S_N(x,y)\bar{\Gamma}] 
      - \Tr[\bar{S}_A(y,x)\Gamma S_A(x,y)\bar{\Gamma}^T].
      \end{split}
    \end{equation}
In this case there is a
disconnected term, and the anomalous term has the opposite sign
to that of the isovector meson. 
From now on the isovector $0^-$
and $1^-$ 
mesons will be referred to as the pion and rho respectively.
Note that for $j=\mu=0$ the difference between $I=0$ and $I=1$ is 
entirely due to disconnected diagrams, which must therefore account for 
the $\pi$ - $\eta^\prime$ mass splitting. For $\mu,\,j\not=0$ this need not be
the case.

In this Letter we for the first time use lattice techniques to study the
spectrum of kaons in a dense baryonic medium. We model the 
kaon states by isovector mesons (\ref{isovector_meson}) in which  
one of the quark flavours does not feel the chemical potential
(for details of how this is achieved see Sec.~\ref{sec:methods}). The resulting
propagator thus approximates the behaviour of an $s$ quark in a medium made up
from $u$ and $d$ quarks, although at this exploratory stage we make no attempt 
to give the strange quarks a realistic mass but simply assume $\kappa_s=\kappa$.
The states studied here 
correspond to $K^\pm$ ($J^P=0^-$) and $K^{*\pm}$ ($J^P=1^-$).

\subsection{Diquarks}

An isoscalar diquark is given by
\begin{equation}
\label{diquark_generic}
D^0(x)=\frac{1}{\surd{2}}\left(\psi_1^T (x)K\bar{\Gamma} 
\psi_2(x)-\psi_2^T(x)K\bar{\Gamma}\psi_1(x)\right)
\end{equation}
implying a correlator:
\begin{equation}
     \label{isoscalar_diquark}
\begin{split}
     2\bra D^{0\dag}(x)D^0(y)\ket &= \bra\psibar_1(x)\Gamma K\psibar_2^T(x)
\psi_1^T(y)K\bar{\Gamma}\psi_2(y)\ket 
     -\bra\psibar_1(x)\Gamma K\psibar_2^T(x)
\psi_2^T(y)K\bar{\Gamma}\psi_1(y)\ket\\ 
      &-\bra\psibar_2(x)\Gamma K\psibar_1^T(x)
\psi_1^T(y)K\bar{\Gamma}\psi_2(y)\ket
      +\bra\psibar_2(x)\Gamma K\psibar_1^T(x)
\psi_2^T(y)K\bar{\Gamma}\psi_1(y)\ket\\
      &=4\Tr[\bar{S}_A(x,x)\Gamma K]\Tr[S_A(y,y)K\bar{\Gamma}]\\ 
      &+2\Tr[S_N(y,x)\Gamma K\bar{S}_N(x,y)\bar{\Gamma}^TK]
      +2\Tr[S_N(y,x)\Gamma K\bar{S}_N(x,y)K\bar{\Gamma}].
\end{split}   
\end{equation}
The last two terms arising from connected diagrams cancel 
if $K\bar{\Gamma}^TK=\bar{\Gamma}$, which is the case for 
$\Gamma = i\gamma_5\gamma_j$, and add up if 
$K\bar{\Gamma}^TK=-\bar{\Gamma}$ implying 
$\Gamma\in\{\One,\gamma_5,\gamma_j\}$. 
Thus the only baryon states available in the $I=0$ sector are $0^+$, $0^-$
and $1^-$;
this is just the Pauli exclusion principle in action.
The isovector diquark correlator is
 \begin{equation}
     \label{isovector_diquark}
     \bra D^{1\dag}(x)D^1(y)\ket 
      =\Tr[S_N(y,x)\Gamma K\bar{S}_N(x,y)\bar{\Gamma}^TK]
      -\Tr[S_N(y,x)\Gamma K\bar{S}_N(x,y)K\bar{\Gamma}].
 \end{equation}
Here the disconnected diagrams cancel, and the connected terms only add up 
for $\Gamma=i\gamma_5\gamma_j$ implying $J^P=1^+$.

Spontaneous symmetry breaking of the U(1)$_B$ symmetry $\psi\mapsto
e^{i\alpha}\psi$, $\bar\psi\mapsto\bar\psi e^{-i\alpha}$ in the superfluid phase
(the symmetry is explicitly broken when $j\not=0$)
can be probed by studying two special diquark states in the isoscalar 
$0^+$ sector which
we refer to as ``Higgs'' and ``Goldstone'' \cite{Hands:2001aq}.
The operators, which by construction yield correlation functions
symmetric under Euclidean time reversal, are given by
\begin{equation}
\label{higgs_prewick}
D^\pm(x)=\psibar_1(x)K\psibar_2^T(x)\pm\psi_1^T(x)K\psi_2(x),
\end{equation}
where the Higgs is `$+$' and the Goldstone `$-$'. 
After Wick contraction the correlators are 
\begin{equation}
\label{higgs}
\begin{split}
\langle D^{\pm}(x)D^{\pm}(y)\rangle &=
\Tr[K\bar{S}_A(x,x)]\Tr[K\bar{S}_A(y,y)]+\Tr[KS_A(x,x)]\Tr[KS_A(y,y)]\\
&\pm\left(\Tr[K\bar{S}_A(x,x)]\Tr[KS_A(y,y)r])+\Tr[KS_A(x,x)]
\Tr[K\bar{S}_A(y,y)]\right)\\
&-\left(\Tr[\bar{S}_A(y,x)K\bar{S}_A(x,y)K]+\Tr[S_A(y,x)KS_A(x,y)K]\right)\\
&\pm\left(\Tr[{S}_N(y,x)K\bar{S}_N(x,y)K]
+\Tr[\bar{S}_N(y,x)KS_N(x,y)K]\right).
\end{split}
\end{equation}
With the choice of diquark source specified in (\ref{eq:action}),
in the superfluid phase the U(1)$_B$ symmetry is preferentially broken by a
condensate $\langle D^+\rangle\not=0$, and as the name implies, in the 
$j\to0$ limit a massless Goldstone mode is interpolated by $D^-$. The
degeneracy between Higgs and Goldstone is principally broken by connected
diagrams formed from normal propagators.

\section{Numerical Method}
\label{sec:methods}

We studied an ensemble of gauge configurations 
generated on a $8^3\times16$ lattice 
at various values of $\mu$
using a Hybrid Monte Carlo algorithm with
the fermion action (\ref{eq:action}) supplemented by a standard Wilson gauge
action  \cite{Hands:2006ve}. 
The parameters were $\beta=1.7$, $\kappa=0.178$; studies of the string
tension suggest a ``physical'' lattice spacing $a=0.26(1)$fm,\footnote{This
value is based on string tension measurements on a $12^3\times24$ lattice and
supplants those reported in \cite{Hands:2006ve, Hands:2006ec}.} 
and studies of the 
$\mu=0$ meson spectrum yield $M_\pi a=0.79(1)$ and $M_\pi/M_\rho=0.80(1)$ 
\cite{Hands:2006ec}. 
For the most part the diquark source $ja=0.04$, though for a few values of 
$\mu$ the series $ja=0.06,0.04,0.02$ was studied in order to permit a $j\to0$
extrapolation.
The values of $\mu$ studied are given in
Table~\ref{table_conf}. Configurations are separated by 4 HMC trajectories of
typical mean
length 0.5, and are not in general independent.
\begin{table}[htbp!]
\centering
\begin{tabular}{|ccc|ccc|}
\hline
$\mu a$ & $ja$ & No. of Configurations&
$\mu a$ & $ja$ & No. of Configurations\\
\hline
0.00 & 0.04 & 210 & 0.50 & 0.02 & 133\\
0.10 & 0.04 & 200 & 0.50 & 0.04 & 248\\
0.25 & 0.04 & 117 & 0.50 & 0.06 & 76\\
0.30 & 0.02 &  78 & 0.55 & 0.04 & 410\\
0.30 & 0.04 & 133 & 0.60 & 0.04 & 441\\
0.30 & 0.06 &  70 & 0.70 & 0.02 & 178\\
0.35 & 0.04 & 124 & 0.70 & 0.04 & 382\\
0.40 & 0.04 & 175 & 0.70 & 0.06 & 70 \\
0.45 & 0.04 & 281 & 0.90 & 0.04 & 65 \\
\hline
\end{tabular}
\caption{Number of configuations examined at 
each value of chemical potential.}
\label{table_conf}
\end{table}

Quark propagators on each configuration were calculated by
all-to-all techniques \cite{Foley:2005ac} 
using time, spin and flavour dilution. 
Whilst this equates to constructing $16\times4\times2=128$ 
dilution vectors per propagator 
and thus performing 256 inversions per configuration, at $\mu\not=0$
this was required for acceptable statistical precision.  
These were saved to disk and stitched into correlation functions 
according to the formul\ae\/ (\ref{isovector_meson}), (\ref{isoscalar_meson}). 
(\ref{isoscalar_diquark}), (\ref{isovector_diquark}) and  (\ref{higgs}).
Disconnected diagrams relevant for each state were calculated and 
saved seperately; 
with the current level of statistics these contributions are both noisy and 
compatable with zero and so only the connected pieces are presented 
in this initial study.

To construct kaon correlators, partial quenching was used 
for the strange propagators, i.e. a further set of propagators 
was calculated on a $\mu\not=0$ gauge configuration using 
${\cal M}(\mu=0;j=0.04)$. This choice of $j$ ensures degeneracy of eg. $K$ and
pion states at $\mu=0$.
The two different sets of propagators were then stitched together with
the relevant operators to form $K$ and $K^*$ states. 
We reiterate that the $s$ quark mass used was identical to that of the 
$u/d$, and clearly whilst the kaon states contain contributions
from virtual $u/d$ loops, 
there is no $s$-quark sea contribution at this stage.

To extract masses the meson correlation functions were fitted to
a cosh function. The time range was adjusted to achieve a stable fit while
minimising the 
obtained $\chi^2/N_{\rm df}$. 
%Sliding window fits and effective mass plots were used in order to
%ensure the correct range 
%was chosen.
For $\mu\not=0$ states with baryon number $B\not=0$
such as the diquarks and the kaons are no longer degenerate 
with their anti-particles. This results in correlators 
%(see fig \ref{graph_diquark_corr}) 
which are no longer 
time-symmetric and 
must be fitted by a sum of two independent exponentials. 
This can lead to problems with fitting as a 4 parameter fit is 
more susceptible to noise than one with just 2. %
%
%\begin{figure}[!hbtp]
%    \centering
%    \includegraphics[width=13.5cm]{diquark_corr.eps}
%    %\vspace{-6mm}
%    \caption{Isoscalar scalar diquark correlators for $\mu a=0.0-0.5$.}
%   \label{graph_diquark_corr}
%\end{figure}
%
Na\"\i vely in the vacuum phase below onset the states' masses 
receive an additive contribution
$\pm\mu BN_c$,  
and so as $\mu$ increases the correlation function 
is increasingly dominated by the lighter of the particle-antiparticle pair, 
which is  
the one with $B<0$. 
For $\mu>0$ the 
heavier state rapidly becomes very difficult to fit.

\section{Results}

All results presented here come purely from analysis of 
hadron correlators formed from connected quark propagators.
The result for the pion mass at $\mu=0$ is $M_\pi a=0.800(3)$.  Chiral
perturbation theory ($\chi$PT)~\cite{Kogut:2000ek}, which models the
weakly-interacting dilute Bose gas forming at onset, can be used to predict the
onset chemical potential $\mu_o$ where the transition from vacuum to the
superfluid phase occurs. This prediction, which strictly speaking only applies
when there is a separation of scales between the pion and heavier hadrons, is
that $\mu_o=M_\pi/2$. In our case, this suggests that the onset in the $j\to0$
limit occurs at $\mu_oa\simeq0.4$.  As $\mu$ is increased we found that more
iterations are required to perform an inversion of the quark matrix
\cite{Hands:2006ve}, especially once $\mu>\mu_o$ due to the large density of
small eigenvalues of ${\cal M}$ in the neighbourhood of the origin in this
regime.

\begin{figure}[!hbtp]
    \centering
    \includegraphics[width=13.0cm]{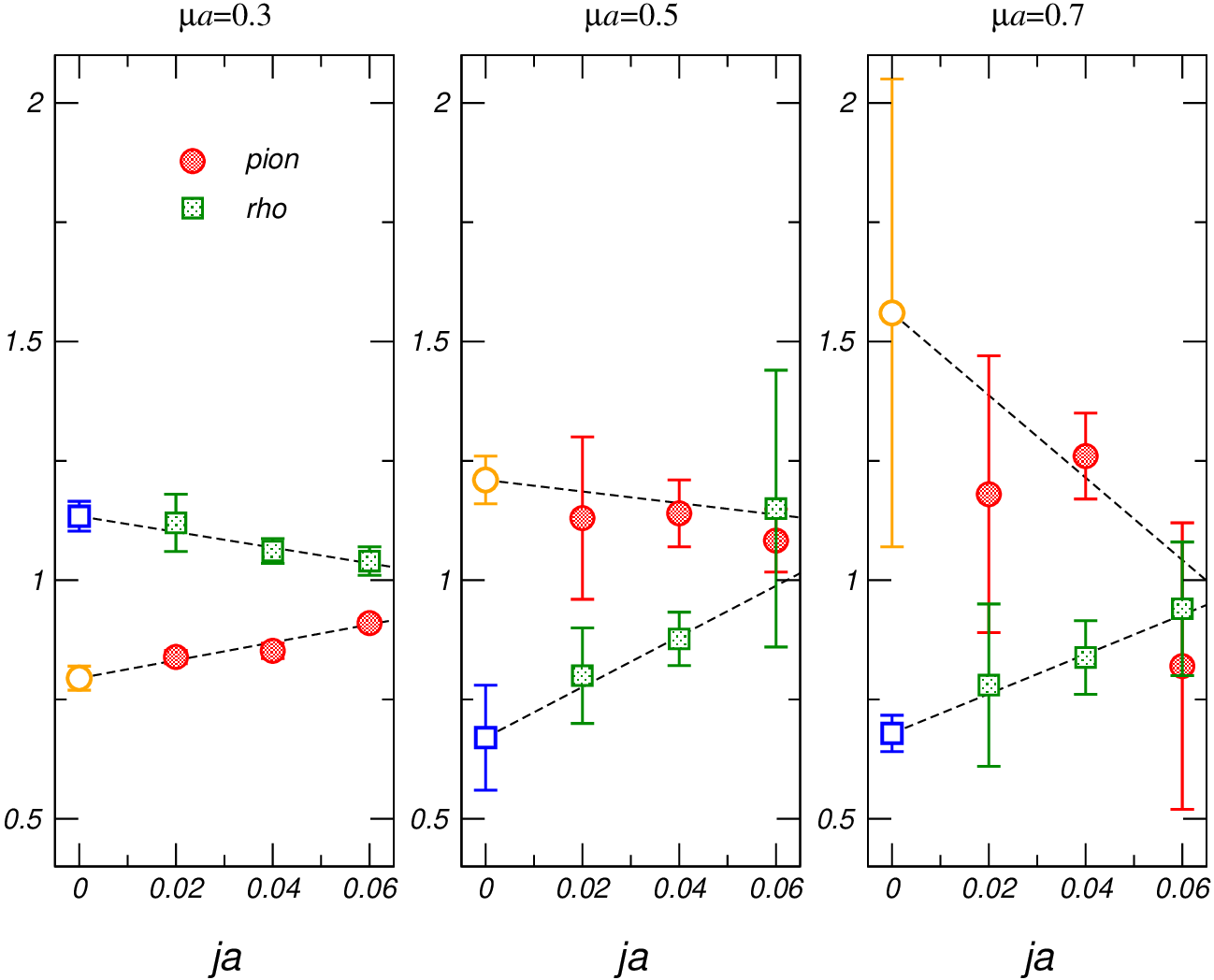}
    %\vspace{-6mm}
    \caption{Pi and rho masses as a function of diquark source
strength $j$ for three representative values of $\mu$, together with their
values
extrapolated to $j=0$.}
   \label{graph_jscaling}
\end{figure}

In principle all our results should be extrapolated to the ``physical'' limit
$j\to0$, but unfortunately available resources preclude a systematic study 
for all $\mu$. Here we follow \cite{Hands:2006ve} by studying
$ja=0.02,\ldots,0.06$ at three representative $\mu$ points: just below onset,
just above onset, and well into the superfluid phase. The results for 
$M_\pi$ and $M_\rho$ are shown in Fig.~\ref{graph_jscaling}. Within the limits
of statistical precision a linear extrapolation $j\to0$ is valid. It is notable
that the pi and rho masses have opposite slopes as functions of $j$,
and that the slopes change sign across the
onset transition between $\mu a=0.3$ and $\mu a=0.5$.

\begin{figure}[!hbtp]
    \centering
    \includegraphics[width=13.0cm]{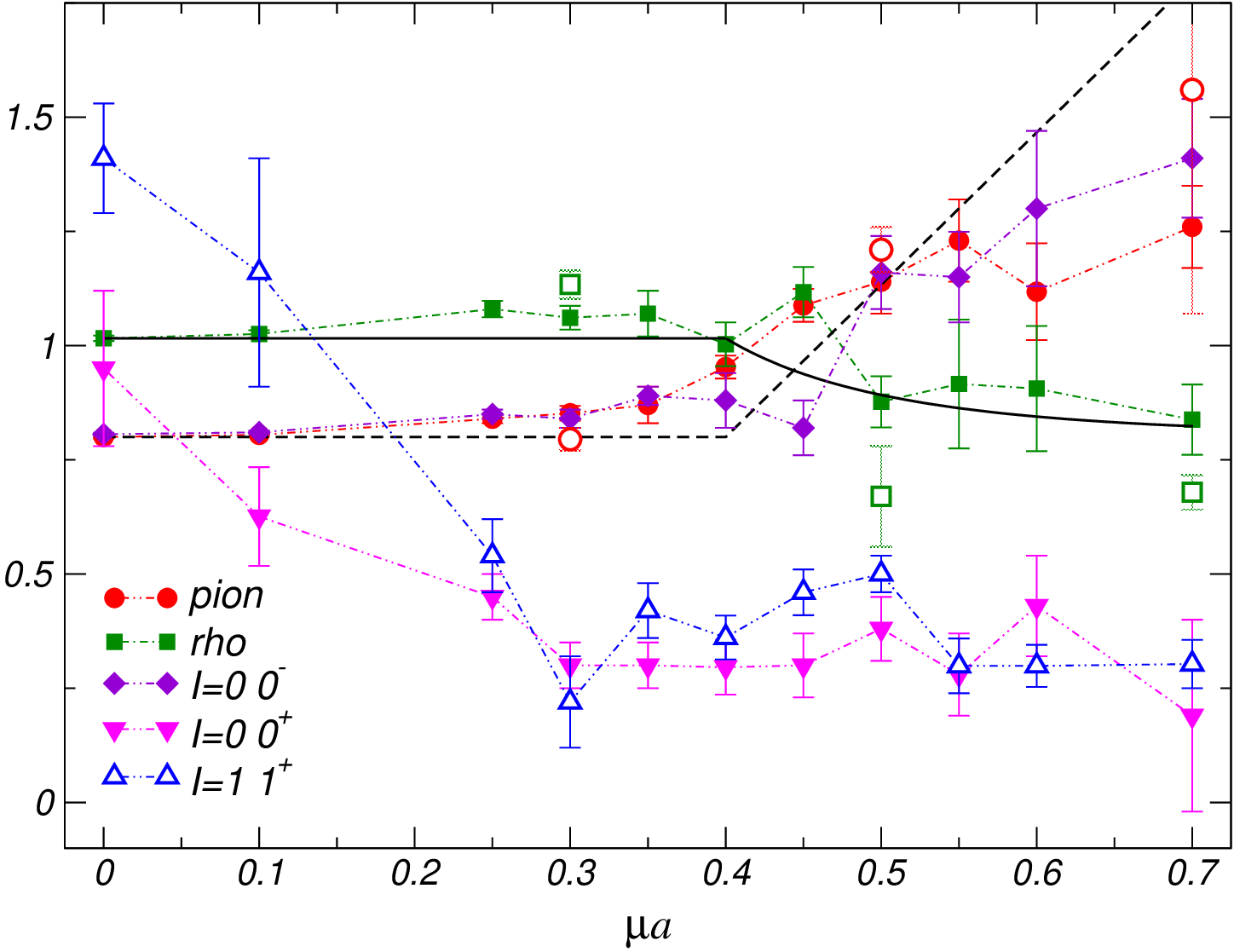}
    %\vspace{-6mm}
    \caption{Meson masses as a function of $\mu$ for $ja=0.04$. 
The dotted line 
is the $\chi$PT prediction for the pion and the solid line is the prediction 
for the rho from \cite{Lenaghan:2001sd}. Open symbols denote extrapolations
to $j\to0$.}
   \label{graph_meson}
\end{figure}

Results for the meson spectrum as a function of $\mu$ are
displayed in Fig.~\ref{graph_meson}. 
At $\mu=0$ the isovector $0^-$ and $1^-$ states, ie. the pion and rho, are
consistent with the values found in \cite{Hands:2006ve,Hands:2006ec}. 
Throughout 
the vacuum phase (i.e. $\mu<\mu_o$) $M_\pi$ and $M_\rho$ are more or less 
constant as expected for states with $B=0$, 
although they do increase slightly but significantly for $\mu a>0.2$. 
For the pion this appears to be a $j\not=0$ artifact vanishing in the $j\to0$
limit, but things are not so clear for the noisier rho.
At and beyond onset at $\mu_o a\simeq0.4$
the pion and rho signals become much noiser as reflected in the error bars,
but it is still possible to identify trends. The pion starts to become 
heavier at onset and appears to increase in mass monotonically with $\mu$ in the
limit $j\to0$.
The $j=0$ $\chi$PT prediction  
$M_\pi=M_\pi(\mu=0)\theta(\mu_0-\mu)+2\mu\theta(\mu-\mu_o)$ \cite{Kogut:2000ek}
(dotted line) is 
followed in a qualitative sense. 
The increase of the pion
mass post-onset is characteristic of a state formed from $q$ and $\bar q$ with a
symmetric combination of quantum numbers under the residual global symmetries
(i.e. the $P_S$ state in the notation of \cite{Kogut:2000ek}) in a theory with 
Dyson index $\beta_D=1$. The same trend is seen in QC$_2$D simulations with 
staggered fermions in the {\em adjoint} representation \cite{Hands:2000ei}, but
the opposite, namely a decrease in $M_\pi$ post-onset, is found in a theory with
$\beta_D=4$ 
such as QC$_2$D with fundamental staggered quarks \cite{Kogut:2001na}.

By contrast post-onset the rho becomes significantly lighter,
in agreement with the result found in simulations on $4^3\times8$ with
significantly heavier quarks by the Hiroshima group \cite{Muroya:2002ry}.
This effect becomes stronger as $j\to0$.
Reduction of $M_\rho$ in a nuclear medium has been proposed to explain the 
low mass
lepton pair enhancement observed in heavy ion collisions \cite{Lenkeit:1999xu}.
With the exception of the
point at $\mu a=0.45$ $M_\rho$ seems to follow the predictions of an effective
model description of spin-1 excitations in QC$_2$D \cite{Lenaghan:2001sd},
shown as a solid line in Fig.~\ref{graph_meson}. 
Our data suggest, however, that the
transition is sharper than that predicted by the model.

\begin{figure}[!hbtp]
    \centering
    \includegraphics[width=12.0cm]{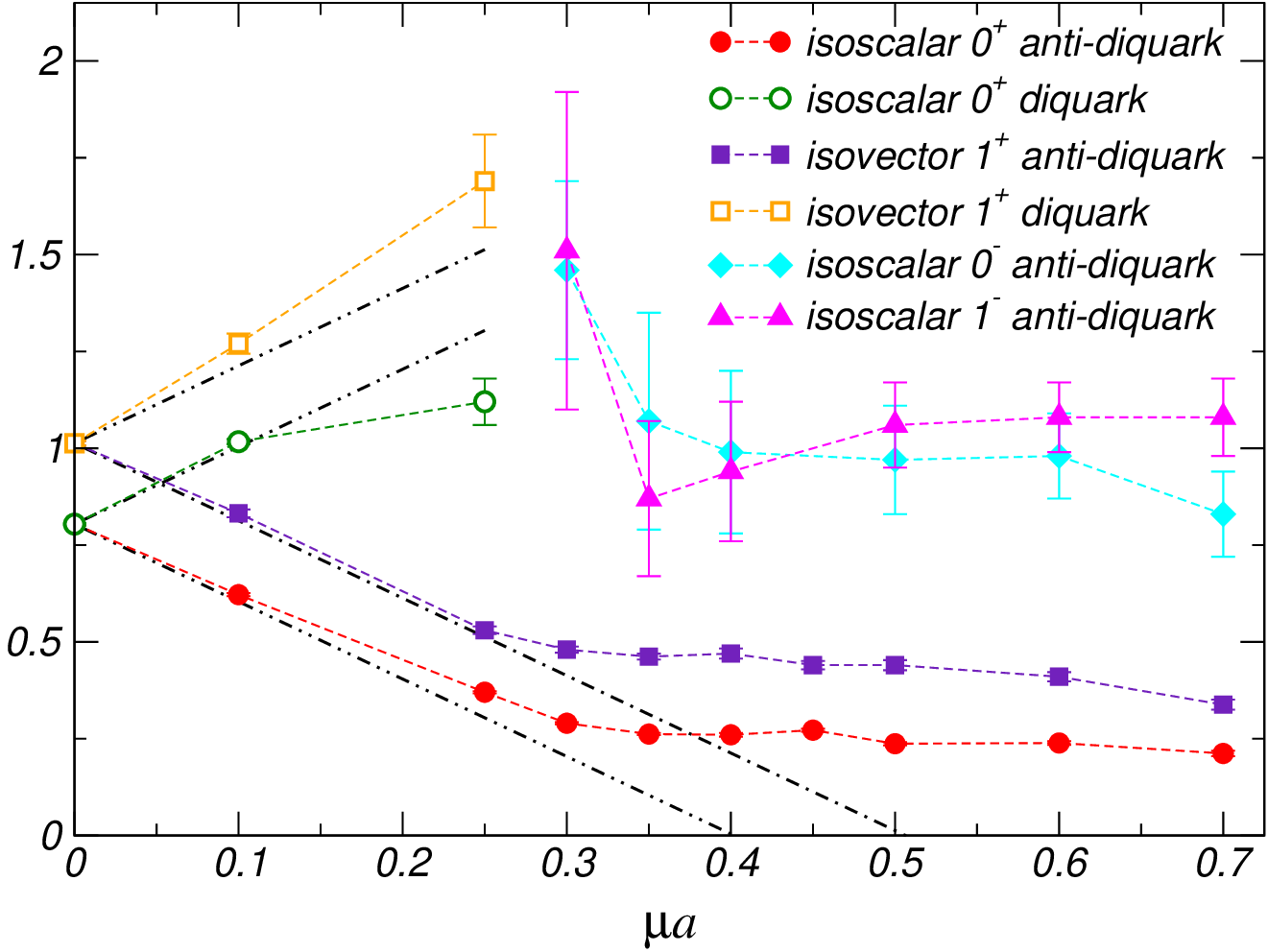}
    %\vspace{-6mm}
    \caption{Diquark 
masses as a function of $\mu$. 
Only states corresponding to operators with
$B<0$ are displayed for 
$\mu a>0.25$. The dot-dashed lines have intercepts at $M_{\pi,\rho}$, 
and gradients $\pm 2$.}
   \label{graph_diquark}
\end{figure}

Results for isovector $0^+$ and isoscalar $1^+$ states are omitted from
Fig.~\ref{graph_meson} because the data are too noisy to fit, and for the
isoscalar 
$1^-$ for the sake of clarity.
The latter follows the rho almost exactly 
until $\mu a>0.5$ at which point it becomes too noisy to measure.
We do, however, include results for the isoscalar $0^-$.
Since only the connected diagrams are currently being 
considered the difference between these two states and the pion and rho is just
the sign of the anomalous term in (\ref{isovector_meson},\ref{isoscalar_meson}),
The fact that they are the same until onset 
shows that this term is 
negligible in this regime.  Beyond onset, there is a small window in which the 
$0^-$ in the isoscalar channel is significantly lighter than the isovector, and
is indeed roughly degenerate with the $I=0$ $0^-$ diquark of
Fig.~\ref{graph_diquark}, to be discussed below.

The two remaining mesons shown in Fig.~\ref{graph_meson}, the isovector 
$1^+$ and isoscalar $0^+$, show a similar behaviour, both starting off 
relatively heavy (and noisy) and then rapidly dropping as $\mu$ increases. 
By $\mu a=0.3$ they have reached a minimum and stay more or less constant 
as $\mu$ increases further. We shall argue below that the low mass of the $0^+$
state is due to its 
overlap with the Goldstone boson in the superfluid phase, when
baryon number ceases to be a good quantum number; the low mass of the $1^+$ is
more unexpected.

To understand the physics of the diquark sector, it is helpful to
begin with the Higgs and Goldstone states of Eq.~(\ref{higgs}) 
with a varying diquark source strength $j$.
\begin{figure}[!hbtp]
    \centering
    \includegraphics[width=14cm]{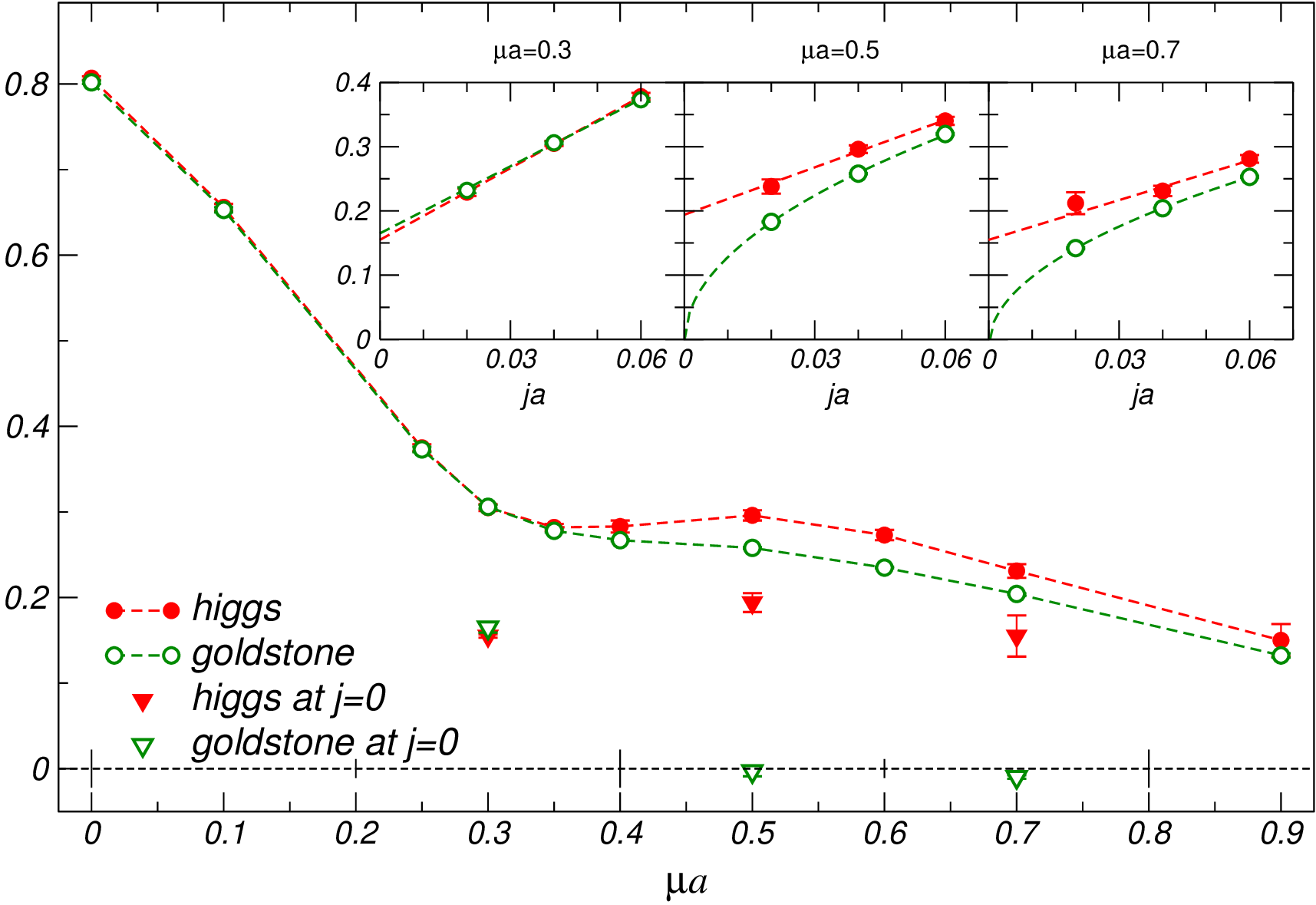}
    %\vspace{-6mm}
    \caption{Higgs and Goldstone masses as a function of 
$\mu$. The two insets show results as $j$ is varied
at fixed $\mu a=0.3,0.5,0.7$. Extrapolations to $j=0$ 
are displayed on the main graph with triangle symbols. }
   \label{graph_higgs}
\end{figure}
Fig.~\ref{graph_higgs} shows Higgs and Goldstone masses as functions of $\mu$
at $ja=0.04$. The insets show how the two states scale with $j$ at three
selected values of $\mu$. 
Below onset Higgs and
Goldstone are degenerate, both scaling approximately
linearly 
\footnote{The pre-onset behaviour
$M(j)=M(0)(1+bj^2)^\frac{1}{4}$ predicted by $\chi$PT \cite{Kogut:2000ek}
may be difficult to distingush from linear behaviour in this regime.} 
with $j$.
Post onset the
degeneracy is broken, and the relation $M_{Gold}\propto\sqrt{j}$
predicted in $\chi$PT \cite{Kogut:2000ek}
appears to hold, the $\chi^2/N_{\rm df}$ 
being 0.67, 0.62 and 1.65 for the 3 linear Higgs extrapolations, 
and 1.08, 2.69 and 0.24 for the 3 Goldstone extrapolations, 
which have the form $M_{Gold}=a\surd j+b$ above onset.

Extrapolations to $j=0$ using these scaling assumptions are also displayed 
on the main graph (triangular symbols).  
The two states remain 
degenerate until onset at which point the Goldstone becomes 
lighter than the Higgs, and appears to become massless as $j\to 0$. 
This is a clear manifestation of spontaneous breaking of U(1)$_B$ symmetry
breaking for $\mu>\mu_o$, implying a superfluid ground state in which baryon
number is no longer a good quantum number, and therefore meson and diquark
states in principle indistinguishable. 

The diquark spectrum in the remaining spin-0 and spin-1 channels 
is shown in Fig.~\ref{graph_diquark}.
It is striking that the signal-noise ratio is much higher for some
diquarks than for the
mesons, also seen in simulations with staggered fermions
\cite{Kogut:2001na}.
The two cleanest signals are for the isoscalar $0^+$ and the isovector $1^+$.
The first observation is that as a consequence of the
symmetry (\ref{eq:MDsymmetry}) there is a relation between the meson and diquark
spectra which holds for $\mu=j=0$ if disconnected diagrams are neglected:
\begin{equation}
M_D(J^P)=M_M(J^{-P}).
\end{equation}
For $0<\mu<\mu_o$, during which the physical ground state remains the vacuum, 
we thus predict $M_D(0^+)=M_\pi\pm2\mu$, $M_D(1^+)=M_\rho\pm2\mu$, 
shown as dot-dashed lines in Fig.~\ref{graph_diquark}.
Indeed both diquark particle-antiparticle pairs behave as expected 
up to $\mu a\approx0.3$. 
Diquark masses
are not shown beyond $\mu a=0.25$ as they become unfittable, as explained in
Sec.~\ref{sec:methods}. 
After this both $0^+$ and $1^+$ anti-diquark states
flatten off and slowly decrease with $\mu$.
The other two isoscalar diquarks constructed from local operators, namely the 
$0^-$ and $1^-$, are extremely heavy and hard to fit below onset, but above
onset have a sufficiently good signal for us to deduce masses 
comparable with $M_\pi(\mu=0), M_\rho(\mu=0)$. 
Although the noise in the 
meson sector is admittedly large, the approximate degeneracy between meson and
baryon sectors in the 
$0^+$ and $1^+$ channels seen in
Figs.~\ref{graph_meson} and \ref{graph_diquark} is consistent with the
meson-baryon degeneracy in the superfluid state discussed above.
Meson-diquark degeneracy has also been observed in quenched studies
at $\mu\not=0$ with staggered fermions~\cite{Giudice:2007yv}.

The isoscalar $0^-$ diquark is 
a particularly interesting state in QC$_2$D, since because of meson-baryon
mixing in the superfluid phase it has the same quantum numbers as 
the $\eta^\prime$ meson \cite{Schafer:2002ty}, and hence its mass
acts as a probe of instanton effects and/or possible restoration of U(1)$_A$
symmetry in a baryonic medium. Unfortunately the current
simulations are not close enough to the chiral limit to settle this
issue via observation of a $\pi(\mu=0)$-$\eta^\prime(\mu)$ mass splitting. 

\begin{figure}[!hbtp]
    \centering
    \includegraphics[width=12.0cm]{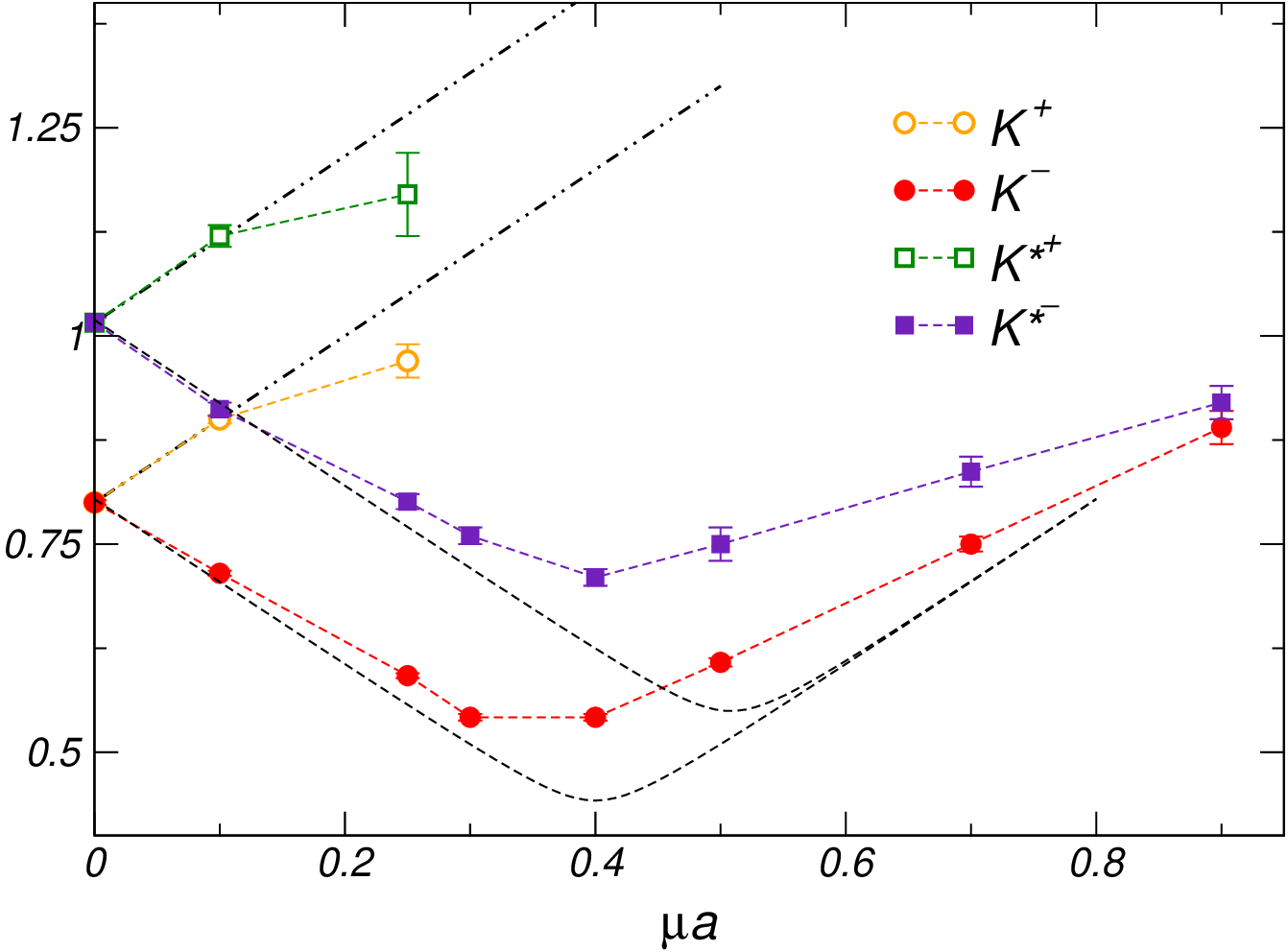}
    %\vspace{-6mm}
    \caption{Kaon spectrum as a function of $\mu$. Only the $K^-$ states are 
displayed for $\mu a>0.25$. 
Dot-dashed lines correspond to $M_K=M_{\pi,\rho}(\mu=0)+\mu$, while
the dashed line corresponds to (\ref{eq:kaon}) with $\vec p=\vec0$ and
$m=M_{\pi,\rho}(\mu=0)/2$}
   \label{graph_kaon}
\end{figure}
The kaon spectrum is shown in Fig.~\ref{graph_kaon}: since $\mu_s$ is assumed to
be zero these states in effect carry a baryon charge conjugate to $\mu$, 
and hence
in general their correlators are not forwards-backwards symmetric. 
Some care must be taken in
assigning physical quantum numbers to the states \cite{Hands:2001aq}. 
Consider say the $0^-$
excitation in a simple model in
which the quarks are non-interacting but have mass $m$. 
Below onset the ``light''
quark spectrum is split into two branches, a hole branch with energy 
$E_h=\sqrt{p^2+m^2}-\mu$ and a particle branch with $E_p=\sqrt{p^2+m^2}+\mu$, 
while the $s$-quark has energy $E_s=\sqrt{p^2+m^2}$ independent of $\mu$. 
We thus identify 
a zero momentum ``$K^-$'' 
state with energy $E_s+E_h=2m-\mu$,
and a ``$K^+$'' with $E_s+E_p=2m+\mu$. 
Above onset, the particle branch becomes filled up to some momentum
$p_F=\sqrt{\mu^2-m^2}$, so that the light quark hole state now costs 
energy $E_h=\mu-\sqrt{p^2+m^2}$. The $K^-$ state now has
$E=E_s+E_h=\mu$ independent of $\vec p$, implying a very small in-medium
velocity. If the discussion is modified to include the effect of a
non-zero diquark source, then the behaviour both below and above onset is
described by the single solution
\begin{equation} 
E(\vec p)=\sqrt{(\mu-\sqrt{p^2+m^2})^2+j^2}+
\sqrt{p^2+m^2+j^2}.
\label{eq:kaon}
\end{equation}

Fig.~\ref{graph_kaon} shows that this simple picture works qualitatively quite 
well in both $K$ and $K^*$ channels, even once interactions are present, 
if we replace the vacuum mass $2m$ with the $\mu=0$ 
mass of the corresponding meson
from Fig.~\ref{graph_meson}. The observed $M_K$ always exceeds the model
prediction, but appears to match both at $\mu=0$ and $\mu\to\infty$ and retains 
a minimum for $\mu\approx\mu_o$. There is thus a significant range
$\mu_o<\mu\lapprox M_K$ in which $M_{K^-}$
lies below its vacuum value, and similarly for $M_{K^{*-}}$.
This offers theoretical support for the idea that
in-medium modification of kaon masses
in nuclear matter 
leads to deeply-bound $K^-$-nuclear states \cite{Akaishi:2002bg}, for which
there is some experimental support from scattering kaons
on light nuclei
\cite{Suzuki:2004ep}. Of course, simulations with realistic
quark masses are needed before this observation can be taken seriously, not to
mention the extension from two colours to three. Nonetheless, we can identify a
minimal requirement for the existence of kaonic-nuclear bound states, namely
$M_K>\mu_o$. This inequality is satisfied in QCD.

\section{Discussion}

\begin{table}[htbp!]
\centering
\begin{tabular}{|cc|cc|cc|}
\hline
Quantum Numbers && \hskip1.5cmMeson&&\hskip1.5cm Diquark&\\
&&$\mu<\mu_o$&$\mu>\mu_o$&$\mu<\mu_o$&$\mu>\mu_o$\\
\hline
$I=0$&$0^+$& noisy& good&good&good\\
$I=0$&$0^-$& good& noisy&noisy&good\\
$I=0$&$1^-$& good& noisy&noisy&good\\
$I=0$&$1^+$& no signal&no signal&\hskip1.5cmno state &\\
\hline
$I=1$&$0^+$& no signal&no signal&\hskip1.5cm no state & \\
$I=1$&$0^-$& good& noisy & \hskip1.5cmno state & \\
$I=1$&$1^-$& good& noisy & \hskip1.5cmno state & \\
$I=1$&$1^+$& noisy& good & good & good \\
\hline
\end{tabular}
\caption{Informal summary of the success of our hadron fits.}
\label{table_summary}
\end{table}
It is helpful to summarise our findings in the form of a table in which the 
quality of our fits both below and above onset is given. The main achievements
of this study have been observations of:

\begin{itemize}

\item
The reversal of the pion and rho levels on crossing from vacuum
into a baryonic medium. In the vacuum $\mu<\mu_o$ $M_{\pi,\rho}$ is
approximately constant, probably because there is no diquark state 
with the same quantum numbers with which to mix.

\item
The breaking of the degeneracy between Higgs and Goldstone diquark 
states
for $\mu>\mu_o$, the Goldstone mass scaling as $\sqrt{j}$ in accordance with 
general theoretical properties of spontaneous symmetry breaking by condensation 
of fermion pairs.

\item
Further evidence for meson-baryon mixing 
in the degeneracy of $I=0$ $0^+$ and $I=1$ $1^+$ states for $\mu>\mu_o$. 
Post onset the $1^+$ appears
to be the next lightest state after the Goldstone and Higgs.
The fact that the mesons with these quantum numbers appear not to have
constant mass even pre-onset (see Fig.~\ref{graph_meson})
can also be ascribed to meson-baryon mixing, since
for $j\not=0$ there is a non-zero amplitude for $\psibar\psi$ to project onto a
baryon.

\item
Evidence of the possibility of bound kaons in nuclear matter.

\end{itemize}

Our study has uncovered little about the effects of a second deconfining
phase transition suspected to occur at $\mu_da\approx0.65$ on this system
\cite{Hands:2006ve}. The only possible discernable trend is a levelling off 
of the already massive pion state for $\mu a\gapprox0.5$ seen in
Fig.~\ref{graph_meson}. Statistical noise increasing with $\mu$, however,
makes this observation provisional at best. In a deconfined phase we might
expect mesons and baryons to be formed from particle-hole and particle-particle 
pairs in the neighbourhood of a Fermi surface, and it is possible that the local
operators used in this study have a poor projection onto the true quasiparticle
excitations. We hope that studies of meson and diquark wavefunctions currently
in progress, as well as extended  operators for the $I=0$ $0^+$ and $I=1$ $1^+$
mesons and enhanced statistics for the disconnected diagrams,
will clarify the situation.

\section{Acknowledgements}

JIS has been supported by IRCSET award SC/03/393Y and SFI grant 04/BRG/P0266.
The project was enabled with the assistance of IBM Deep Computing.
We'd also like to thank Justin Foley for helpful discussions.

\bibliography{ref}

\end{document}